# THE SLUGGS SURVEY: *HST*/ACS MOSAIC IMAGING OF THE NGC 3115 GLOBULAR CLUSTER SYSTEM

Zachary G. Jennings[1], Jay Strader[2], Aaron J. Romanowsky[1,3], Jean P. Brodie[1], Jacob A. Arnold[1], Dacheng Lin[4], Jimmy A. Irwin[4], Gregory R. Sivakoff[5], Ka-Wah Wong[4]
*Draft version April 13, 2018*

## ABSTRACT

We present *Hubble Space Telescope*/Advanced Camera for Surveys (*HST*/ACS) $g$ and $z$ photometry and half-light radii $R_h$ measurements of 360 globular cluster (GC) candidates around the nearby S0 galaxy NGC 3115. We also include Subaru/Suprime-Cam $g$, $r$, and $i$ photometry of 421 additional candidates. The well-established color bimodality of the GC system is obvious in the *HST*/ACS photometry. We find evidence for a "blue tilt" in the blue GC subpopulation, wherein the GCs in the blue subpopulation get redder as luminosity increases, indicative of a mass-metallicity relationship. We find a color gradient in both the red and blue subpopulations, with each group of clusters becoming bluer at larger distances from NGC 3115. The gradient is of similar strength in both subpopulations, but is monotonic and more significant for the blue clusters. On average, the blue clusters have $\sim 10\%$ larger $R_h$ than the red clusters. This average difference is less than is typically observed for early-type galaxies but does match that measured in the literature for the Sombrero Galaxy (M104), suggesting that morphology and inclination may affect the measured size difference between the red and blue clusters. However, the scatter on the $R_h$ measurements is large. We also identify 31 clusters more extended than typical GCs, which we consider ultra-compact dwarf (UCD) candidates. Many of these objects are actually considerably fainter than typical UCDs. While it is likely that a significant number will be background contaminants, six of these UCD candidates are spectroscopically confirmed as NGC 3115 members. To explore the prevalence of low-mass X-ray binaries in the GC system, we match our ACS and Suprime-Cam detections to corresponding *Chandra* X-ray sources. We identify 45 X-ray - GC matches, 16 among the blue subpopulation and 29 among the red subpopulation. These X-ray/GC coincidence fractions are larger than is typical for most GC systems, probably due to the increased depth of the X-ray data compared to previous studies of GC systems.

*Subject headings:* galaxies:individual:(NGC 3115) - galaxies: star clusters: general - globular clusters: general

## 1. INTRODUCTION

Globular clusters (GCs) serve a valuable role in the study of extragalactic systems. Due to their old ages, the properties of GCs trace the earliest stages of galaxy formation (see the review by Brodie & Strader 2006). In addition, due to their high luminosities, they are more easily observable than faint galaxy starlight, allowing for detailed inferences of galaxy formation and evolution in nearby systems at large galactocentric distances.

Extensive multi-wavelength investigation of nearby early-type galaxies has revealed a number of interesting properties in their GC systems. It has been well established for decades that early-type GC systems display clear color bimodality (e.g., Zepf & Ashman 1993; Ostrov et al. 1993). It is generally accepted that this color bimodality corresponds to an underlying metallicity bimodality, with the red clusters metal-enhanced when compared to the blue clusters. This metallicity bimodality has been spectroscopically confirmed for a limited, but growing, number of systems (Puzia et al. 2002; Strader et al. 2007; Beasley et al. 2008; Alves-Brito et al. 2011; Usher et al. 2012; Brodie et al. 2012).

In addition, it is well established that a disproportionate number of low-mass X-ray binaries (LMXBs) in early-type systems are found in GCs (Fabbiano 2006). The host GCs tend to be the most dense and compact clusters, where dynamical interactions are capable of creating LMXB systems. The exact properties of GCs that host these LMXBs are not well understood. It is generally established that the metal-rich subpopulation contains a significantly greater number of LMXBs, but it is not known how these properties may depend on other galaxy environmental factors (e.g., Sivakoff et al. 2007; Kundu et al. 2007).

In this paper, we investigate the GC system of the nearby S0 galaxy NGC 3115 with a six pointing *Hubble Space Telescope*/Advanced Camera for Surveys (*HST*/ACS) mosaic in the F475W and F850LP filters (hereafter $g$ and $z$). *HST* observations of nearby GC systems have critical advantages over ground based imaging. The high resolution afforded by *HST* means that GCs will be partially resolved, allowing for measurements of their half-light radii ($R_h$). Extensive studies of partially-resolved GCs in *HST* imaging have revealed that red clusters are typically smaller than blue clusters, an observation that may be explained by either projection effects (Larsen & Brodie 2003) or intrinsic differences (Jordán 2004a).

However, while *HST* observations are powerful due to their resolution, they also feature a limited field of view. Ground based imaging is generally able to probe the halos of nearby galaxies out to many effective radii ($R_e$), but *HST* requires multiple mosaiced images to achieve similar radial coverage. In general, while numerous studies have in-vestigated GC properties in ACS imaging (e.g., Jordán et al. 2005, 2007a), there has been only limited exploration of GC trends out to several $R_e$ (e.g. Spitler et al. 2006; Forbes et al. 2006; Nantais et al. 2011; Blom et al. 2012; Strader et al. 2012; Usher et al. 2013; Puzia et al. 2014). In particular, there have been few studies of the GC systems of lenticular galaxies in *HST*/ACS imaging, especially in non-cluster environments (e.g. Spitler et al. 2006; Cantiello et al. 2007; Harris & Zaritsky 2009; Harris et al. 2010; Forbes et al. 2010). Finally, cluster size information has the added benefit of allowing us to search for ultra compact dwarf (UCD) candidates in our images.

[1] University of California Observatories, Santa Cruz, CA 95064, USA; zgjennin@ucsc.edu
[2] Department of Physics and Astronomy, Michigan State University, East Lansing, Michigan, MI 48824, USA
[3] Department of Physics and Astronomy, San José State University, One Washington Square, San Jose, CA, 95192, USA
[4] Department of Physics and Astronomy, University of Alabama, Box 870324, Tuscaloosa, AL 35487, USA
[5] Department of Physics, University of Alberta, Edmonton, Alberta, T6G 2E1, Canada

NGC 3115 is of particular interest due both to its proximity and the properties of its GC system, and we are studying it as part of the SAGES Legacy Unifying Globulars and GalaxieS survey (SLUGGS; Brodie et al. 2014[6]). It is highly inclined and located at $D = 9.4$ Mpc (Tonry et al. 2001, with recommended correction of $(m-M) = -0.06$ from Mei et al. 2007). Both the galaxy as a whole, as well as its GCs, have seen previous discussion in the literature. Elson (1997) identified a color bimodality in the stellar halo of NGC 3115 in *HST*/WFPC2 data. Kundu & Whitmore (1998) subsequently analyzed the GCs seen in the WFPC2 data and identified a color bimodality in this population as well.

The GC system has also been analyzed spectroscopically. Puzia et al. (2002) identified a metallicity bimodality in the GC population using VLT/ISAAC spectroscopically, and found both populations consistent with being coeval, albeit with large uncertainties. Kuntschner et al. (2002) confirmed the GC bimodality in VLT/FORS2 spectroscopy of 24 GCs and found evidence for multiple formation epochs in the GC system. Brodie et al. (2012) demonstrated the metallicity bimodality to high confidence using a sample of 71 GCs with CaT-derived metallicities from DEIMOS spectra. Norris et al. (2006) found kinematic links between the NGC 3115 stellar spheroid component and the red GC population, as well as overall rotation in the GC population. Arnold et al. (2011) considered combined GC and stellar spectra and found that the properties of both the GC system and overall galaxy light favor a distinct 2-phase formation scenario wherein the halo of the galaxy is built through a series of minor mergers. The overall inferences from previous studies are that the color and metallicity bimodality in the GC systems are unambiguous, and that the GC system is clearly evolutionarily linked to the build up of the overall galaxy light. If there are intrinsic differences between the red and blue GC populations other properties (i.e. size or X-ray frequency), the NGC 3115 GC population is an ideal place to look.

For the remainder of this work, we adopt the $R_e = 57''$ value for the bulge of NGC 3115 from Capaccioli et al. (1987) for consistency with Arnold et al. (2011), equivalent to $\approx 2.6$ kpc projected distance. We also adopt the flattening value of $q = 0.5$ from Arnold et al. (2011) and a heliocentric recessional velocity of 663 km s$^{-1}$ from Norris et al. (2006).

In §2, we discuss our methodology, particularly our methods for carrying out photometry and $R_h$ measurements in our ACS images. In §3, we discuss our findings, including confirmation of color bimodality in the ACS data, the discovery of a "blue tilt," trends in color and $R_h$ with galactocentric distance, and the results of a search for UCD candidates in our sample. In §4, we match our ACS catalog with *Chandra* X-ray detections and identify clusters with associated X-ray emission.

Finally, in §5 we summarize our results.

## 2. DATA ANALYSIS

In this section, we discuss the methods employed to create our GC catalogs, consisting primarily of photometry and $R_h$ measurements in our ACS mosaic. We also supplement our analysis with spectral catalogs of NGC 3115 GCs from Arnold et al. (2011) and Pota et al. (2013), as well as a catalog of Subaru/Suprime-Cam *g*, *r*, and *i* photometry of the GC system initially analyzed in Arnold et al. (2011) but not fully published. We include the full catalog in this paper for reference.

### 2.1. *Initial Data Reduction*

The primary dataset analyzed in this work is the ACS/WFC mosaic of NGC 3115 from *HST* Program 12759 (PI: Jimmy Irwin). The mosaic consists of six pointings of the galaxy, extending out to $\sim 5R_e$. Exposures are 824 s in *g* and 1170 s in *z*, with the exception of the POS-3 pointing (labelled 3 in Fig. 1), which was observed with exposures of 722 s in *g* and 1137 s in *z*. A simple line dither was used to cover the ACS chip gap, with 2 total exposures in *g* and 3 in *z* at each pointing. The two central pointings overlap significantly in the center in order to increase the signal-to-noise ratio (S/N) for the innermost GCs, where galaxy light adds additional noise. In Fig. 1, we overlay our six ACS pointings on the combined *g*, *r*, and *i* Subaru/Suprime-Cam mosaic.

We downloaded the *.flc* files of the exposures from the MAST website. These files have been flat-fielded and corrected for charge transfer efficiency problems in the ACS instrument. We then used the *astrodrizzle*[7] package to drizzle the separate images into distortion-corrected mosaics for each filter at each pointing. We performed a conservative CR rejection during the drizzling process to reject obvious cosmic rays.

### 2.2. *Initial Photometry Measurements*

We created our initial catalog for analysis using SExtractor (Bertin & Arnouts 1996), selecting all candidates that were $3\sigma$ above background. We also required that each selection have at least 10 connected pixels. Photometric zeropoints were calculated using the ACS *PHOTPLAM* and *PHOTFLAM* keywords, placing the magnitudes on to the AB system. It is worth noting that while the *HST*/ACS filters are very close to the SDSS filters, they do not have precisely the same response functions, and slight systematic offsets are not unexpected.

We performed aperture photometry on this full list using the *daophot* package in *IRAF*. We measured magnitudes within a 5 pixel (0.25$''$) aperture to maximize S/N, which we corrected to a 10 pixel (0.5$''$) aperture based on average photometric measurements of bright GCs in our data. The values of these corrections are 0.175 in *g* and 0.249 in *z*. Finally, we corrected these magnitudes to pseudo-infinite apertures (5.5$''$) using the values from Sirianni et al. (2005), who measured corrections of 0.095 is *g* and 0.117 in *z*. We also corrected for galactic foreground reddening using the Schlegel et al. (1998) reddening maps with updated calibrations from Schlafly & Finkbeiner (2011, $A_V = 0.130$ for NGC 3115).

For the largest GCs in our sample, the above aperture corrections will systematically underestimate the final magnitudes due to additional light falling outside the 10-pixel corrected aperture. To correct for this, we employed size-dependent aperture corrections as described in §2.4.1. Based on the consistency of our photometry with matched Subaru/Suprime-Cam photometry (see §2.5), we conclude that our photometric methods are reasonable. It is worth noting that the size-dependent corrections are less well-constrained for the largest clusters due to the low S/N in the wings of the ACS PSF. As a result, while the size-dependent corrections are an appropriate first-order correction for the luminosities of the largest clusters, there will still be lingering systematic uncertainties.

---

[6] http://sluggs.ucolick.org

[7] http://www.stsci.edu/hst/HST_overview/drizzlepac

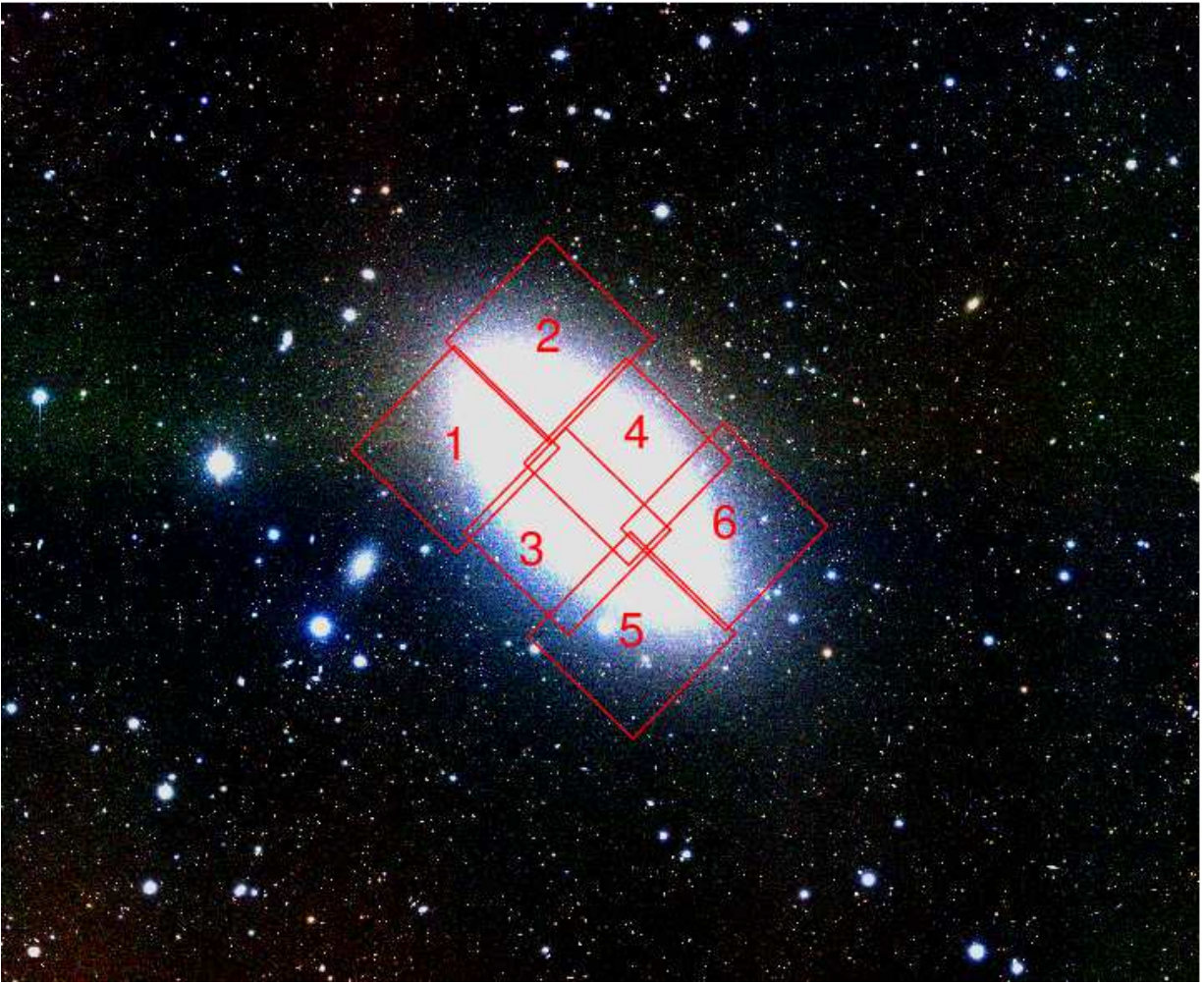

**Figure 1.** Combined *g*, *r*, and *i* Suprime-Cam image of NGC 3115. *g* is colored blue, *r* green, and *i* red. The locations of our six *HST*/ACS pointings are overlaid. North is up and east is left. Each box approximately represents the ACS field of view and is 200″ on a side.

None of our results depend strongly on the actual measured luminosities of the GCs, and systematic errors in the aperture corrections of GCs have little effect on the measured color of our sources. The latter point has also been emphasized in previous ACS studies of extragalactic GC systems (Jordán et al. 2009; Strader et al. 2012).

After photometry was performed, we pruned our catalog down to a list of reasonable GC candidates. First, we required that each object was detected in both filters by matching the coordinate lists within 2 pixels. We also rejected objects fainter than $g = 26$ or $z = 25$ to remove spurious noise detections and objects significantly fainter than we would expect for the GC population (typical turn-over magnitudes for GC populations will be $g \sim 22.5$ at the distance of NGC 3115, see Jordán et al. 2007b). Several of our pointings feature overlap, especially across the center of NGC 3115. We measured a simple astrometric shift to transform all pointings to the same WCS. We then matched sources across both pointings to within 3 pixels to combine photometry from sources imaged in multiple pointings. For all such sources, our final magnitudes are a weighted average of the separate photometric measurements. The median offset between sources detected in multiple images is 0.023 mag.

### 2.3. *ishape $R_h$ Measurements*

The principal advantage of ACS GC imaging is the superior angular resolution of the instrument, which allows us to resolve the angular size of globular clusters. Throughout the paper, we quantify this size in terms of a half-light radius ($R_h$) and use the terms interchangeably. At the distance of NGC 3115, GCs of typical radius ($\sim$ 2–4 pc) will be partially resolved. The point spread function (PSF) of a partially resolved cluster is a convolution of the intrinsic PSF of the ACS instrument and the light profile for GCs, which we assume to be described by a King profile. We measured the PSF empirically using bright, unsaturated stars in the ACS field. The PSF was measured separately for each filter, but the same PSF was used across all pointings for each filter. We then used *ishape* (Larsen 1999) to measure the FWHM of the King profile of the GC, which is easily converted to a half-light radius $R_h$ using the relation $R_h = 1.48(\text{FWHM}_{King30})$ (Larsen 1999). The cluster concentration c (defined as $c \equiv r_t/r_0$, where $r_t$ is the tidal radius and $r_0$ is the core radius) cannot be measured reliably for most clusters; as a result, we adopted a King profile with a fixed value of $c = 30$ for all cluster measurements, consistent with other partially resolved GC studies (e.g., Harris & Zaritsky 2009; Strader et al. 2012).

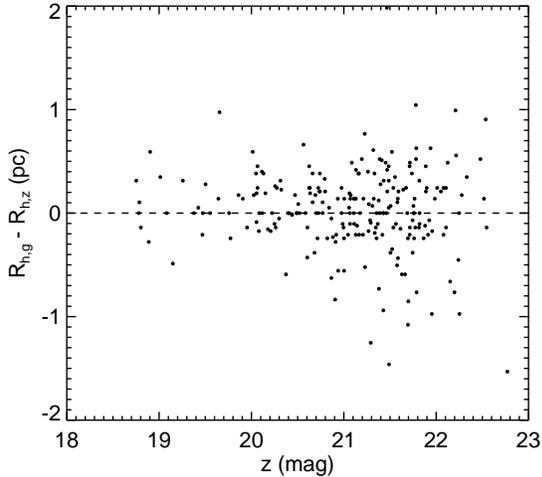

**Figure 2.** Plot of difference in $R_h$ as measured in each of our two filters, $g$ and $z$. We find a median offset of $(R_{h,g} - R_{h,z})_{med} = 0.00$ pc. The standard deviation of the differences is 0.55 pc. We interpret this as good agreement between the two filters, and use the weighted average of the two individual $R_h$ measurements for the remainder of this work.

We made measurements of $R_h$ using both filters. In Fig. 2, we plot the difference between sizes measured in both filters, $R_{h,g} - R_{h,z}$. We measure precisely zero median offset between the filter measurements. The standard deviation of the differences is 0.55 pc, indicating good agreement between the filters with some scatter in $R_h$ measurements for the same source. In general, the scatter of the difference is larger both for more extended clusters and for fainter sources. Note that some clusters have exactly the same measurement in multiple filters due to internal resolution limits in ishape. For the remainder of the paper, we adopt the weighted average of the $R_h$ value measured in the two filters. Previous studies of ACS-measured GC sizes have found systematic uncertainties on the level of ∼20%, or ∼0.4 pc for a typical cluster size of 2.0 pc, when measuring sizes with *ishape* (Spitler et al. 2006; Harris & Zaritsky 2009).

Harris (2009) argued that measurements of $R_h$ are only reliable for candidates with S/N above 50, which roughly corresponds to limits of $g \sim 24$ and $z \sim 23$ in our data. While we measure sizes for all candidates in our sample, when we examine trends in the sizes of the GC subpopulations, we only consider those clusters with S/N above 50, around 65% of our sample.

### 2.3.1. *Size-Dependent Aperture Corrections*

To correct for additional light outside the 10 pixel aperture, we employed size-dependent aperture corrections for the full sample using a similar method to Strader et al. (2012). We convolved the empirically measured PSF from the ACS imaging with King profiles of fixed concentration 30, the same as used for the *ishape* measurements. We varied the FWHM of the King profile to create a series of fake clusters from $R_h = 0$ pc out to $R_h = 40$ pc. Using aperture photometry, we measured the light excess outside the 10 pixel aperture as a function of input cluster size, which we then used to correct our measured photometry. Representative corrections in $g$ are 0.004 mag for $R_h = 2$ pc, 0.30 mag for $R_h = 10$ pc, and 0.94 mag for $R_h = 30$ pc. These values are larger than those in Strader et al. (2012) because NGC 3115 is ∼ 7 Mpc closer than NGC 4649. As a result, clusters of similar size occupy a much larger angular area in the NGC 3115 data, requiring larger aperture corrections.

We performed corrections separately for the $g$ and $z$ filters, using the empirically measured PSF for both. However, since the excess light for large clusters is dominated by the King profile, the influence of the empirical PSF is minimal and thus the difference in the corrections between the two filters is small. As a result, the colors of the clusters are essentially unchanged by these corrections, as discussed in §2.2.

Note that uncertainties in the measured sizes of clusters are neglected in these photometric measurements. For GC-sized objects, these uncertainties are negligible given the small magnitudes of the corrections. However, for the larger objects in our sample, uncertainties in the measured sizes of objects can be large, especially for objects which are not actually well-parameterized by King profiles (i.e. background galaxies).

### 2.4. *Catalog Selection*

We performed *ishape*[8] (Larsen 1999) measurements on the remaining ACS detected sources, as explained in §2.4. We considered any source with a measured size less than 0.3 pc a likely point source and removed it from the final GC candidate catalog. A color cut was applied to the catalog, selecting sources between $0.5 \leq (g-z) \leq 1.7$. We also performed a by-eye rejection of obvious background galaxies with visible features. Finally, we used our measured *ishape* sizes to remove extended objects, as described in §2.4.2. After all cuts, we were left with 360 GC candidates in our final catalog.

Table 1 summarizes the number of GC candidates in our various catalogs. We also list the number of GCs in the red and blue subpopulations, the division of which is explained in §3.1.

### 2.4.1. *Use of $R_h$ Measurements in Catalog Selection*

Our use of *ishape* size measurements in pruning our final catalog merits further discussion. In any catalog cuts we may consider, we must make a trade-off between rejecting as much contamination as possible while preserving bona fide clusters. This motivates the use of different cuts in size for different scientific measurements of interest. We investigate three different catalogs in §3.

First, for investigation of trends in cluster color and magnitude, we are chiefly interested in rejecting contaminating foreground stars and background galaxies that happen to fall in our GC color-magnitude space. As a result, we reject any source with a measured $R_h < 0.3$ pc as being a point source, and therefore either a foreground star or background AGN. In addition, any source with a measured $R_h > 8.0$ pc was rejected as being a possible background galaxy. While these cuts may also reject a few actual GCs from our sample, this rejection will simply increase random uncertainties from having fewer clusters. Note that, for these rejections, we still incorporate $R_h$ measurements for clusters with S/N below 50.

---
[8] http://baolab.astroduo.org/

**Table 1**
Summary of GC Catalogs Analyzed in this Work

| Catalog | Total GCs | (Total With Spectral Confirmation) | Blue GCs | (With Spectra) | Red GCs | (With Spectra) |
|---|---|---|---|---|---|---|
| ACS Detected GCs | 360 | (134) | 191 | (64) | 169 | (70) |
| Suprime-Cam Only GC Candidates | 421 | (42) | 299 | (24) | 122 | (18) |
| All GC Candidates | 781 | (176) | 490 | (88) | 291 | (88) |

Note: Since spectroscopic data were acquired before the *HST*/ACS data, no GCs unique to the ACS catalog have spectroscopically measured velocities.

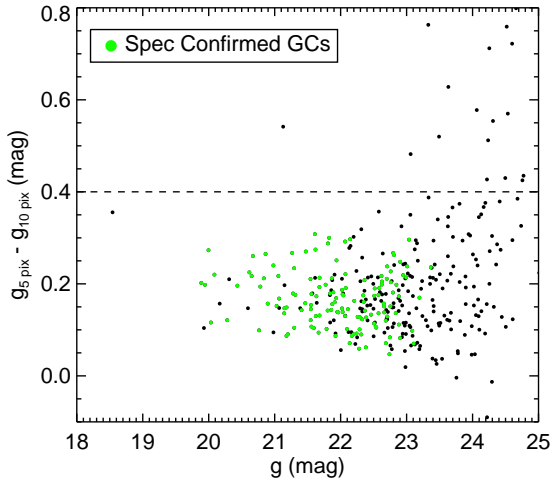

**Figure 3.** Illustration of our aperture-difference galaxy rejection method. We plot the difference in $g$ as measured in 5 and 10 pixel apertures against the full corrected $g$ value. We consider those clusters with $g_{5\mathrm{pix}} - g_{10\mathrm{pix}} > 0.4$ as being extended sources and remove them from our highest confidence GC catalog. Spectroscopically-confirmed GCs are plotted in green.

These are probably not very reliable for the larger sized objects, but there is no obvious selection bias introduced in the colors by their removal.

Many extended sources tend to be in the magnitude range where *ishape* is not as useful for rejection of galaxies. These sources will be extended, but with measured sizes that are still consistent with the largest GCs and luminosities that are consistent with the faint end of the globular cluster luminosity function (GCLF). To deal with such sources, we employ an aperture difference measurement: we reject any object with $g_{5\mathrm{pix}} - g_{10\mathrm{pix}} > 0.4$ as being extended, allowing us to cull down our final candidate list to a reasonable selection of GCs. We plot this selection criterion, which rejected 28 sources, in Fig. 3. We consider this catalog, containing 360 sources, as our final GC candidate list.

For §3.4, dealing with specific $R_h$ measurements, we also reject all objects with a S/N below 50. For this analysis, we are interested in more precise measurements of GC $R_h$, rather than crude rejection of point sources or highly extended objects. This list of high S/N GCs contains 235 objects.

Finally, for §3.5, we are interested in finding ultra compact dwarf (UCD) candidates, as well as other clusters with sizes more extended than typical for GCs. As we are interested in reliably measured sizes for larger objects, we still require that the S/N of these candidates be greater than 50. However, as we are specifically interested in identifying sources with large $R_h$, we remove all constraints on the maximum size of the clusters, including both $R_h$ measurement rejection and aperture difference rejection. This list of large UCD candidates contains 31 sources. Note that this list still discards objects rejected in our by-eye step that have obvious morphological features of galaxies.

### 2.5. *Ground Based Photometry and Spectroscopy*

To provide comparison with earlier work and improve our catalog selection, we supplement our ACS data with ground based imaging and spectroscopy. Arnold et al. (2011) presented a photometric and spectroscopic study of GCs around NGC 3115 using a combination of Subaru $g$, $r$, and $i$ photometry and DEIMOS, LRIS, and IMACS spectroscopy. In this work, we publish the full catalog of Surpime-Cam(Miyazaki et al. 2002) photometry from the Arnold et al. (2011) study. The full spectroscopic sample was subsequently published in Pota et al. (2013), and we include these velocities where available. The catalog includes 176 GCs with measured radial velocities consistent with NGC 3115 membership. In addition, there are 421 point-sources without measured velocities but with $g$, $r$, and $i$ colors consistent with GCs. The color cuts adopted by Arnold et al. (2011) correspond to

$$0.5 \leq (g-i)_0 \leq 1.4 \quad (1)$$

and

$$0.45 \times (g-i)_0 - 0.026 \leq (g-r)_0 \leq 0.45 \times (g-i)_0 - 0.08. \quad (2)$$

The Arnold et al. (2011) catalog also removes objects fainter than $i = 23$.

We display the color–color diagram in Fig. 4, with the entire point source catalog ($\sim 20000$ objects) in black and those which passed the color–color cut in blue. In addition, we plot the location of GCs with ACS measured sizes in red and spectroscopically confirmed GCs in green. It is clear that color-color selections from ground based imaging reject a number of strong GC candidates, which is a necessary trade-off given the contamination of foreground and background sources away from the GC color-color sequence. Note that some spectroscopically confirmed GCs are outside the color-color selection as well. These GCs were placed on DEIMOS slitmasks as "low-confidence" GCs outside the main color selection, but ended up being confirmed regardless, further indicating the incompleteness of any sample selected with simply unresolved photometry.

Note that this catalog likely includes significant contamination from foreground stars and background galaxies, which are difficult to reject in ground-based imaging without measured radial velocities. We matched this GC catalog with our ACS data and identified 187 sources consistent with both catalogs.

We did not use the Suprime-Cam catalog $g$, $r$, and $i$ color–color photometry to reject any ACS objects that otherwise passed our ACS CMD and size cuts. While this would allow us to reduce background galaxy contamination in the ACS catalog, it would also introduce additional selection biases into our sample and reject a number of true GCs. We have greater confidence in ACS identified selections with consistent sizes than in non-spectroscopically confirmed Suprime-Cam color–color selections. However, we still include the measured Suprime-Cam photometry for these cases in our data table for reference.

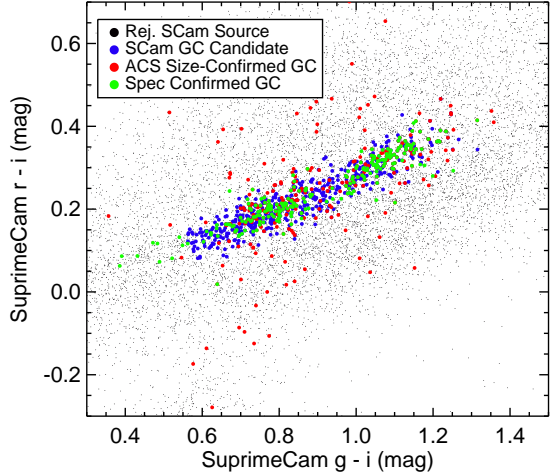

**Figure 4.** SuprimeCam $(g-i)_0$ vs. $(r-i)_0$ color–color diagram. All detected point sources are plotted in black, while those that pass our color–color and FWHM cuts are plotted in blue. We also plot GCs with ACS-measured sizes in red and spectroscopically confirmed GCs in green.

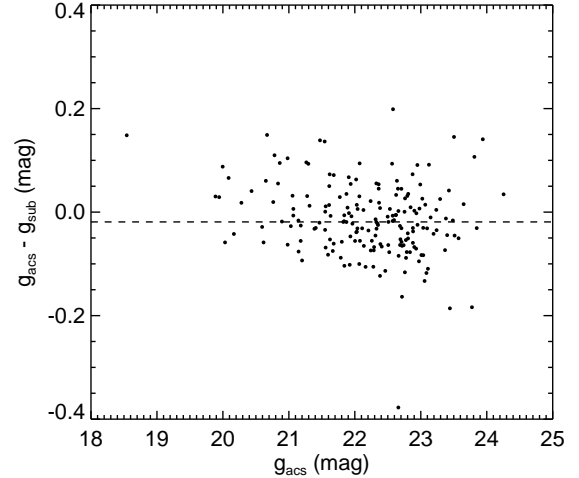

**Figure 5.** Difference between $g_{acs}$ and $g_{sub}$ for GC candidates common to both the ACS catalog and the Subaru Suprime-Cam catalog. The median offset between the photometry is 0.018 mag and the standard deviation of the differences is 0.078 mag, indicating good agreement.

Note that significant selection biases factor into which catalog a source will be found in. The Suprime-Cam catalog will be less complete very close to the nucleus of NGC 3115, and will also not reach as faint as the ACS catalog. In addition, brighter GCs were preferentially selected for spectroscopy due to observational constraints; it is difficult to get reliable spectra for GCs fainter than $i \sim 22$, and essentially impossible for objects fainter than $i \sim 23$. Finally, as the Suprime-Cam catalog was used to select spectral targets, no GCs unique to the ACS catalog have spectral confirmation.

It is worth noting that better selection can be achieved from color-color selections if NIR photometry is incorporated. Muñoz et al. (2014) demonstrated that incorporation of K-band photometry in color-color selections provides a much better discriminant between GCs and contaminants. There are still difficulties with identifying real GCs outside typical color-color selections. However, as more NIR photometric datasets of nearby galaxies become available (e.g. NGVS-IR), contamination of foreground and background objects in purely photometric GC studies can be significantly reduced.

We briefly detail our methods for Subaru/Suprime-Cam photometry below. Methodology for spectroscopic data analysis was presented in Arnold et al. (2011).

### 2.5.1. *Subaru/Suprime-Cam Photometry*

NGC 3115 was imaged in $g$, $r$ and $i$-band filters on January 4th, 2008 using Suprime-Cam on the 8.2-m Subaru telescope. The camera's 34′-by-27′ field-of-view consists of a mosaic of ten 2k-by-4k CCDs, separated by an average gap width of 16″. A series of short exposures were taken with each filter ($g$: 5x40s, $r$: 5x15 s, $i$: 5x15 s) using a pre-defined 5-point dither pattern to account for bad pixels and to fill in chip gaps. Seeing varied between 0.5″ and 0.7″.

Suprime-Cam data were reduced using the *SDFRED1* pipeline[9]. Standard aperture Photometry was performed using IRAF, with zero points computed from SDSS sources in the field of view of a different galaxy observed the same night. This is necessary as NGC 3115 is not in the SDSS footprint. We also performed a color-correction to place our measurements onto the SDSS filter system[10]. Our apertures were selected to maximize the S/N. We then performed an aperture correction using a larger aperture and several bright stars located in the field. Astrometry was calibrated using the USNO-B catalog.

We compared our ACS measured $g$ magnitudes to those from Subaru imaging. In Fig. 5, we plot $g_{acs} - g_{sub}$ against $g_{acs}$. For GC candidates in both catalogs, we found a median difference of $g_{acs} - g_{sub} = -0.018$ with the standard deviation of the differences being 0.078. Thus while there is some scatter in our measured photometry for a given source, in general $g$ magnitudes can be compared without regard for any sizable systematic offset. Since GCs are unresolved point-sources in the Subaru photometry, a size-dependent aperture correction is not applicable to the Subaru imaging.

### 2.5.2. *Use of Radial Velocities in Catalog Selections*

For purposes of our catalog, we rejected any sources with measured radial velocities less than 350 km s$^{-1}$ as being Milky Way foreground stars. While there will likely be a small amount of GCs with radial velocities smaller than this, Milky Way star contamination becomes dominant for velocities below this. For reference, the systematic radial velocity of NGC 3115 is 663 km s$^{-1}$, with a typical rotational velocity of $\sim$240 km s$^{-1}$ and a dispersion of $\sim$100 km s$^{-1}$ (Norris et al. 2006). Naturally, the latter two quantities vary with radius. We did not use radial velocities to reject background objects. Only six of our photometrically selected GCs have measured radial velocities greater than 1000 km s$^{-1}$, the highest of which is 1210 km s$^{-1}$. As there are no objects with velocities drastically inconsistent with NGC 3115 measurement, we included all these objects in our GC catalog. Inclusion of these six objects ultimately makes no difference to our overall conclusions.

### 2.6. *Globular Cluster Luminosity Function*

As a check on the reasonableness of our catalog selection criteria, we plot and fit the measured g-band GCLF for GC

---

[9] http://www.naoj.org/Observing/Instruments/SCam/sdfred/sdfred1.html.en

[10] http://www.sdss.org/dr7/algorithms/jeg_photometric_eq_dr1.html#usno2SDSS

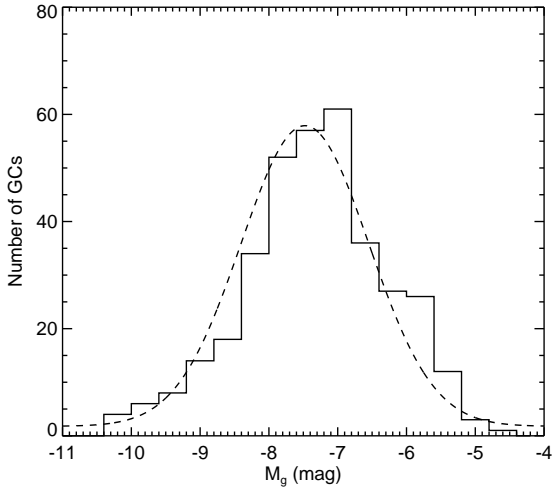

**Figure 6.** Histogram of GC candidate absolute magnitudes, with Gaussian fit plotted. The luminosity function peaks at $M_g = -7.4$, consistent with studies of the GCLF around other early-type galaxies. We take this as evidence that our selection criteria produce a reasonable GC catalog, and that our data are reasonably complete down to the faintest magnitudes at which we would expect to find GCs. The poor fit at low luminosities is indicative of the level at which background contamination sets in (roughly $g \sim 23$ in apparent magnitude).

candidates observed for our ACS sample in Fig 6. We recover a reasonably well-defined GCLF. A Gaussian fit returns a peak located at $M_g = -7.4, \sigma_g = 0.97$, which is consistent with that observed in ACS studies of other early-type galaxies (Jordán et al. 2007b). As we recover reasonable parameters for GCLF, we conclude that our catalog selection criteria produce a well-selected GC candidate list. Note that we have neglected any sort of contamination or completeness correction in the fitting of the GCLF. The ACS mosaic is deep, so it is likely that we are quite complete down to low luminosities. However, the GCLF fits more poorly at lower luminosities. This is an indication of the level at which contamination of the sample by background galaxies sets in, around roughly $g \sim 23$ in apparent magnitude.

It is worth noting that the selection biases in the Suprime-Cam catalog and the ACS catalog are different. In color-color space, background galaxies overlap little with GCs, and as a result the 3-filter Suprime-Cam catalog offers an effective means of background galaxy rejection. However, foreground star contamination is still prevalent in this catalog. Since GCs are partially resolved in ACS images, it is fairly easy to differentiate between unresolved foreground stars and partially resolved GCs in the ACS catalog. However, background galaxy rejection is somewhat ambiguous in the ACS data (although still reasonably achieved with color and size selections).

### 2.7. *X-Ray Observations*

The *HST*/ACS mosaic examined in this study was acquired to provide companion optical photometry to a $\sim 1$ Msec *Chandra* observation of NGC 3115. The full details of the *Chandra* data analysis will be presented in a companion paper (Lin D. et al. 2014, in preparation). In this work, we match the observed X-ray sources from this catalog to GC candidates in the ACS mosaic to find GCs which may harbor LMXBs. To treat uncertainty on the position, we adopt the 3-$\sigma$ uncertainty on the X-ray position source position; we assume that the uncertainties on the ACS-measured positions are small in comparison to the uncertainties on the X-ray-measured positions.

A more thorough investigation of the choice of selection annulus will be included in the companion paper. Typical values of this uncertainty are $\sim 0.4''$. If any ACS GC candidates are found within the 3-$\sigma$ uncertainty of the X-ray detection, we classify the closest ACS GC candidate as having an X-ray match. We did not encounter any situations in the ACS data where multiple GCs were within the 3-$\sigma$ radius.

For Suprime-Cam detections far from the center of the *Chandra* pointing, the *Chandra* PSF is large, and there are a handful of *Chandra* sources where the 3-$\sigma$ uncertainty overlaps with multiple Suprime-Cam sources. As we do not perform any analysis of the Suprime-Cam/X-ray matches in this work, we do not attempt to resolve these ambiguities further in this work and simply associate the single nearest source to the X-ray centroid as matching the X-ray detection. We briefly comment on apparent features of those GCs that hosts X-ray sources, specifically the coincidence fractions, in §4.

### 3. RESULTS AND DISCUSSION

In this section, we examine the photometry and sizes measured for our GC candidates. To explore trends in these values as a function of galactocentric radius, we adopt an equivalent "elliptical radius", defined as

$$R_g = \sqrt{qX^2 + Y^2/q}, \quad (3)$$

where $q = 1 - b/a = 0.5$ is the flattening parameter for NGC 3115 and $X$ and $Y$ are cartesian coordinates measured along the major and minor axes of an ellipse centered on NGC 3115. While magnitudes of various gradients do change slightly (as would be expected) if we instead parameterize our radial trends using a simple circular radius, none of our overall conclusions are changed. Unless otherwise noted, $R_g$ is measured in kiloparsecs for the remainder of the paper. We adopt the same P.A.= $43.5°$ as Arnold et al. (2011). Unless otherwise noted, we quote 1 $\sigma$ uncertainties on any fitted relations.

In a few sections, we will discuss GC masses in place of luminosities. For these conversions, we must naturally use a mass-to-light ratio. We adopt a constant $M/L = 1.45$ from the arguments presented in Sivakoff et al. (2007) for all such conversions in the remainder of the paper.

In Table 2, we list a sample of our full catalog. *HST*/ACS objects are listed first, with objects only detected in Subaru Suprime-Cam imaging following. The entire catalog is available online in a machine readable format.

### 3.1. *Color Bimodality*

In Fig. 7 we present the measured ACS color-magnitude diagram (CMD) of the GC system in NGC 3115. The color bimodality of the NGC 3115 system, while well established, is particularly obvious in the ACS catalog. Using Gaussian Mixture Modelling (Muratov & Gnedin 2010), we find that a unimodal distribution is rejected at greater than 99.9% confidence. It has also been shown that the color bimodality of the system directly corresponds to a metallicity bimodality in the GC subpopulation, with the blue subpopulation being more metal-poor than the red subpopulation (Brodie et al. 2012). The clear division between the red and blue subpopulations implies that differences between the two subpopulations in other properties should be particularly obvious. Using a Gaussian kernel density estimate, we find that the subpopulations are separated at $g - z = 1.13$ (adopting a smoothing kernel of 0.07). We adopt this color as the dividing line between the two subpopulations, color the metal-poor subpopulation blue and

Table 2
ACS and SuprimeCam GC Candidate Catalog

| ID | R.A. (J2000) (deg.) | Dec. (J2000) (deg.) | ACS $g$ (mag) | err[A] (mag) | ACS $z$ (mag) | err[A] (mag) | SCam $g$ (mag) | err (mag) | SCam $r$ (mag) | err (mag) | SCam $i$ (mag) | err (mag) | $R_h$ (pc) | Vel.[B] (km s$^{-1}$) | X-Ray?[C] (Y/N) |
|---|---|---|---|---|---|---|---|---|---|---|---|---|---|---|---|
| A1 | 151.315689 | -7.714322 | 18.542 | 0.001 | 17.232 | 0.001 | 18.394 | 0.016 | 17.632 | 0.016 | 17.375 | 0.014 | 6.07000 | - | N |
| A2 | 151.290795 | -7.720553 | 19.928 | 0.002 | 18.752 | 0.002 | 19.989 | 0.003 | 19.308 | 0.004 | 18.992 | 0.003 | 1.69000 | - | Y |
| A3 | 151.304573 | -7.732728 | 19.944 | 0.002 | 18.78 | 0.002 | 19.915 | 0.003 | 19.222 | 0.004 | 18.941 | 0.004 | 2.44000 | 494 | N |
| A4 | 151.325999 | -7.735499 | 20.087 | 0.002 | 18.786 | 0.002 | 20.021 | 0.003 | 19.268 | 0.004 | 18.968 | 0.003 | 2.49000 | 1123 | Y |
| A5 | 151.333993 | -7.695504 | 20.033 | 0.002 | 18.802 | 0.002 | 20.091 | 0.004 | 19.36 | 0.004 | 19.062 | 0.003 | 1.74000 | 682 | Y |
| A6 | 151.338241 | -7.693624 | 19.888 | 0.002 | 18.888 | 0.002 | 19.858 | 0.003 | 19.221 | 0.004 | 19.012 | 0.003 | 2.99000 | 1028 | N |
| A7 | 151.289973 | -7.700146 | 20.281 | 0.002 | 18.903 | 0.002 | 20.263 | 0.004 | 19.506 | 0.004 | 19.174 | 0.004 | 1.76000 | 697 | Y |
| A8 | 151.304950 | -7.704758 | 20.167 | 0.002 | 19.013 | 0.002 | 20.209 | 0.004 | 19.486 | 0.004 | 19.217 | 0.004 | 2.47000 | - | Y |
| A9 | 151.348540 | -7.698285 | 19.996 | 0.002 | 19.082 | 0.002 | 19.909 | 0.003 | 19.321 | 0.004 | 19.159 | 0.003 | 4.70000 | 736 | N |
| A10 | 151.295165 | -7.728766 | 20.31 | 0.003 | 19.148 | 0.004 | 20.138 | 0.028 | 19.457 | 0.03 | 19.205 | 0.023 | 2.42000 | - | Y |

**A:** Errors on ACS photometry measurements only include statistical uncertainties on the initial photometry, and neglect systematic uncertainties from size-dependent aperture corrections.
**B**: Heliocentric velocity, if available, from the Pota et al. (2013) catalog.
**C**: Whether or not the GC candidate hosts an X-ray source.
Note: The above table is fully available online in machine readable format. A sample is shown here to represent its content.

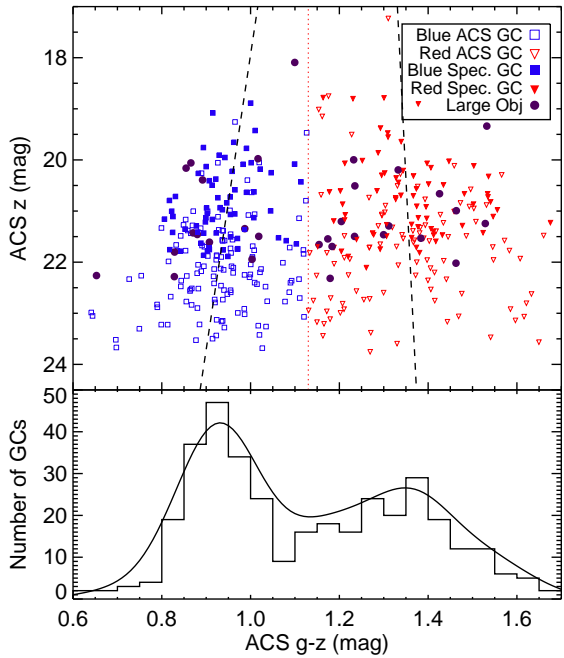

**Figure 7.** Top Panel: Observed CMD of GCs in our catalog, following application of our various quality cuts. Metal-poor and metal-rich subpopulations are plotted in blue and red respectively. Spectroscopically confirmed GCs from Arnold et al. (2011) are plotted as solid symbols, while those without spectroscopic confirmation are plotted as open symbols.. The well-studied bimodality of the GC system is clear in our data. The color dividing line is located at $g-z = 1.13$ and marked with a red dashed line. The "blue-tilt" mass-metallicity relation is clear in the blue subpopulation. There are also hints of an opposite trend in the very brightest metal-rich clusters, but it is of low significance. Bottom Panel: Color histogram of GC candidates, with Gaussian kernel density estimate overplotted. The Gaussian density plot and histogram are scaled to contain the same total number of GCs. The bimodality of the system is clearly visible.

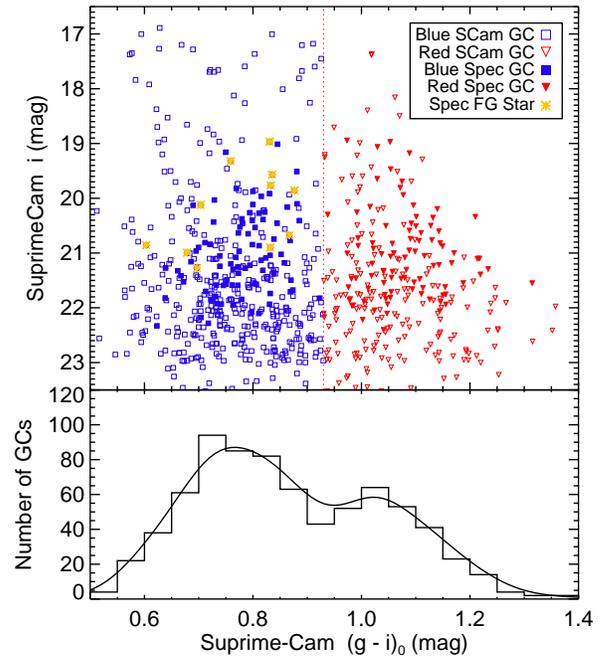

**Figure 8.** Top Panel: CMD of Suprime-Cam detected GCs. GCs with spectroscopic confirmation are plotted as sold symbols, while those without spectroscopic confirmation are plotted as open symbols. Bottom Panel: $(g-i)_0$ histogram for Suprime-Cam imaging, with Gaussian kernel density estimate overplotted. The division between blue and red subpopulations is located at $(g-i)_0 = 0.93$ and is marked with a dashed red line. The Gaussian density estimate is normalized so that the total number of GCs is the same as the histogram.

the metal-rich subpopulation red, and retain this scheme in our subsequent figures. We also plot extended objects $R_h > 8$ pc as purple points. These objects are not included in our fits nor our kernel density estimate.

Many of our candidate GCs have corresponding radial velocity measurements from Arnold et al. (2011). As mentioned in §2, we discard those candidates with measurements which exclude NGC 3115 membership. Those spectroscopically confirmed candidates with correct radial velocities are plotted with solid symbols, while those without spectroscopic confirmation are plotted as open symbols in Fig. 7.

To compare our measured colors with metallicities, we adopt the following conversions from Peng et al. (2006a):

$$[Fe/H] = -6.21 + (5.14 \pm 0.67) \times (g-z) \quad (4)$$

if $0.7 < (g-z) \leq 1.05$, and

$$[Fe/H] = -2.75 + (1.83 \pm 0.23) \times (g-z) \quad (5)$$

if $1.05 < (g-z) < 1.45$.

Using these conversions, the divide between metal-poor and metal-rich GCs is located at $[Fe/H] \sim -0.8$ dex.

For comparison, in Fig. 8, we plot the $(g-i)_0$ CMD from our photometrically-selected Suprime-Cam catalog. Again, bimodality is clear in the data; a Gaussian kernel density estimate places the dividing line between the red and blue subpopulations at $(g-i)_0 = 0.93$. As before, we plot spectroscopically confirmed candidates as solid symbols, while those without spectroscopic confirmation are plotted with open symbols. In addition, for comparison, we include spectroscopically confirmed foreground stars, which were not included in the Gaussian kernel estimate. A similar histogram is also displayed in the lower panel.

It is interesting to note the increased prevalence of blue GC candidates in the Suprime-Cam catalog, compared to the ACS CMD (see also Table 1). This is likely due to a combination of two effects. First, the spatial distribution of blue GC candidates around NGC 3115 is more extended than that of the red GC candidates. As the Suprime-Cam FOV is much larger than the ACS FOV, this extended population is preferentially sampled by the larger Suprime-Cam catalog. However, it is also apparent that most of the spectroscopically confirmed foreground stars are blue. There will certainly be more foreground contaminants in the purely photometrically selected GCs, and this will increase the relative number of blue GCs. Disentangling these two effects is difficult without spectroscopic confirmation, and in reality both effects will contribute to the larger relative number of blue GC candidates in the Suprime-Cam photometric catalog.

In Fig. 9, we plot the locations of all the blue and red GC candidates in our sample. ACS detected GCs are plotted with solid circles, while those with open circles are only detected in the Suprime-Cam imaging. The blue GC subpopulation is clearly more spatially extended than the red subpopulation. We mark the locations of extended ($R_h > 8$ pc) clusters in our sample with X symbols.

### 3.2. *The Blue Tilt*

Blue tilts, wherein blue GCs tend to become increasingly red with increasing brightness, have been observed in many extragalactic GC systems (e.g., Strader et al. 2006; Harris et al. 2006; Mieske et al. 2006b). These gradients are typically taken as evidence of a mass-metallicity relationship among the clusters. This is likely due to self-enrichment; brighter, more massive clusters are able to retain more of their enriched material, producing redder (more metal-rich) photometric measurements. We find evidence for a blue-tilt in the GC system in ACS photometry. A least squares fit to the blue subpopulation gives a relation of

$$(g-z)_{\rm blue} = (-0.017 \pm 0.006) \times z + (1.31 \pm 0.14), \quad (6)$$

which we plot in Fig. 7. If we restrict our fit to only those GCs which are spectroscopically confirmed, we find a slope of $-0.024 \pm 0.012$, indicating that even our most conservative sample still displays evidence for a blue-tilt.

The existence of corresponding red-tilts in extragalactic GC subpopulations is on somewhat shakier observational ground. In our sample, a linear fit to the red subpopulation gives

$$(g-z)_{\rm red} = (0.006 \pm 0.009) \times z + (1.23 \pm 0.20), \quad (7)$$

indicating that the red subpopulation may tilt in the opposite direction (brighter clusters are slightly bluer than dimmer clusters). However, the slope is not significant and is dependent on the degree of rejection of background galaxies, which almost always tend to be contaminants on the red end of the GC subpopulation. Fitting only the spectroscopically confirmed red clusters gives a slope of $0.012 \pm 0.17$, again not significantly constrained.

The strength of the blue tilt we measure for NGC 3115 is actually less than is typically seen for other galaxies (e.g. Strader et al. 2006; Usher et al. 2013). Typical measured slopes from similar ACS filter sets have found blue tilt slopes around $-0.040$ mag mag$^{-1}$ for similar filters. Our measured value is roughly half of this. A plausible explanation for this smaller slope is the proximity of NGC 3115. While our data reach similar limiting apparent magnitudes as other extragalactic ACS GC studies, we are looking a magnitude further down the GCLF compared to other studies. If we convert the measured luminosity to a mass estimate, we are probing GCs of roughly half the mass in our ACS data.

Bailin & Harris (2009) proposed a quantitative self-enrichment model for GCs. The principle result was that GCs above the mass of $\sim 2 \times 10^6$ M$_\odot$ (roughly $z = 19.2$ in our sample) will display a mass–metallicity correlation, while GCs less massive will not. In addition, a weak red tilt in the same direction as the blue tilt was also predicted. We attempted to measure the slope of the blue tilt for only the brightest clusters in our sample. While the slope became more negative as we restricted our sample to brighter and brighter GCs, we found that once we cut out clusters with magnitudes fainter than $z = 22$, the blue tilt detection becomes of marginal significance. For GCs brighter than $z = 21$, the detection is less than 1-$\sigma$. Thus while the magnitude of the blue tilt does appear to approach the slopes seen in previous studies for our brightest GCs, we cannot make this observation with any statistical significance.

Looking for a blue tilt in the photometrically selected Suprime-Cam sample instead gives a tilt in the opposite direction, with brighter candidates appearing bluer, in both color subpopulations. However, as mentioned in §3.1, we believe the purely photometric Suprime-Cam catalog contains significant contamination from foreground stars and does not constitute a sample of pure GCs. Given that the spectroscopic sample still displays a standard blue tilt, we consider this merely indicative of the degree of foreground contamination in the Arnold et al. (2011) sample.

### 3.3. *Trends in Color with Radius*

In Fig.10, we plot the observed color of the GCs in $(g-z)$ against their galactocentric elliptical radius $R_g$ from the center of NGC 3115. For illustrative purposes, we plot median colors for GCs in each subpopulation for 7 equal-numbered bins, binned as a function of distance. The red bins each contain 24 GCs, while the blue bins each contain 27. We also plot 68% uncertainties on these medians, estimated through bootstrapping. Both subpopulations appear to display gradients, with clusters becoming bluer farther out from the center of NGC 3115. To evaluate the color gradients in both subpopulations, we perform a least squares fit to the individual data points for the two subpopulations. We find the following relations:

$$(g-z)_{\rm Red} = (-0.05 \pm 0.04) \times \log R_{\rm g} + (1.38 \pm 0.02) \quad (8)$$

for the red subpopulation and

$$(g-z)_{\rm Blue} = (-0.06 \pm 0.02) \times \log R_{\rm g} + (0.97 \pm 0.02) \quad (9)$$

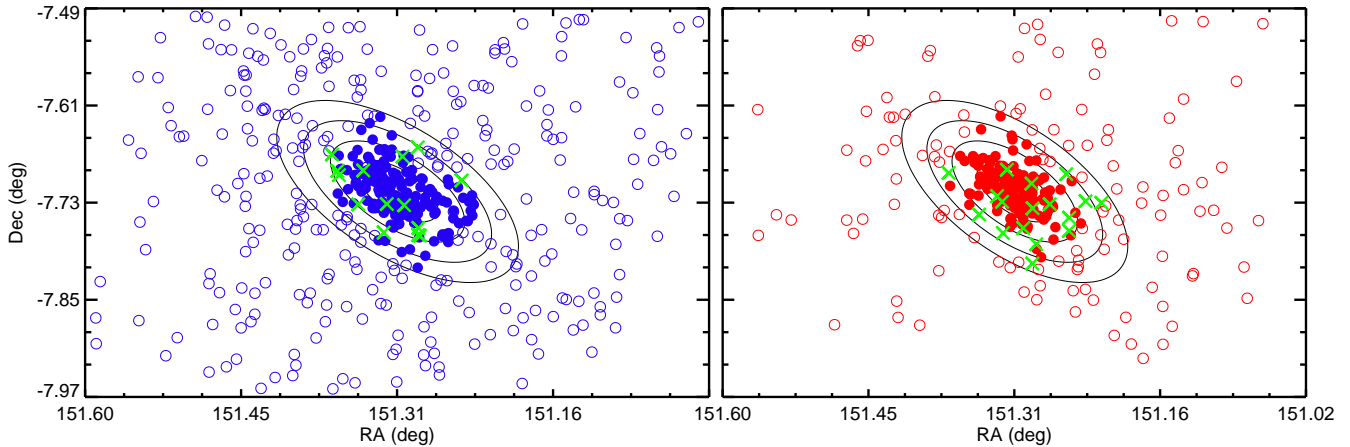

**Figure 9.** Spatial locations of blue and red GCs around NGC 3115. Filled circles are those GCs detected in the ACS imaging, while open circles are those only detected in the Suprime-Cam imaging. We also plot ellipses representing 1, 3, 5, 7, and 9 $R_e$ (1 $R_e \approx 2.6$ kpc) from NGC 3115. The blue GCs tend to be more spatially extended than the red GCs. Extended objects with $R_h \geq 8.0$ pc are marked with green X's.

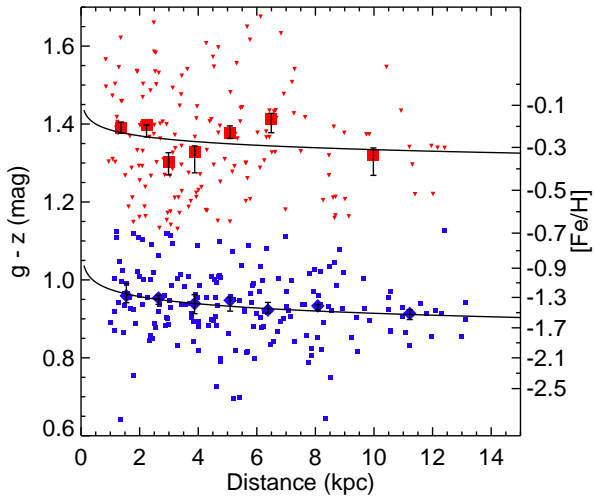

**Figure 10.** Plot of GC color vs. elliptical distance from NGC 3115 for our clusters. Clusters are colored according to their subpopulation. We also plot the median $g-z$ colors of both subpopulations for 7 equal number bins in each subpopulation. 68% uncertainties on the median colors from bootstrapping are also included. Both subpopulations display a color gradient, with clusters becoming bluer as they get farther away from NGC 3115. The blue clusters decrease uniformly, while the red clusters appear to display visible substructure in color as a function of distance. Least squares fits to both subpopulations are also plotted.

for the blue subpopulation. These relations are plotted in Fig 10.

We also compute simple linear gradients from least squares fitting, giving $(g-z)_{\rm Red} \propto (-0.004 \pm 0.003) R_g$ and $(g-z)_{\rm Blue} \propto (-0.005 \pm 0.002) R_g$. We convert our measured colors to metallicities using the Peng et al. (2006a) relations above and estimate metallicity gradients in [Fe/H] of $-0.10 \pm 0.07$ dex dex$^{-1}$ for the red subpopulation and $-0.29 \pm 0.11$ dex dex$^{-1}$ for the blue subpopulation, neglecting uncertainties on the conversion factors. The two gradients are clearly very comparable in color, while the [Fe/H] gradient is stronger for the blue subpopulation due to the relationship between color and metallicitiy for GCs. The decrease in color for the blue subpopulation is essentially monotonic, while the medians in the red subpopulation instead display some visible substructure, a trend seen in radial color studies in other GC systems (Strader et al. 2012). However, the uncertainties on the median quantities for the red subpopulation are large, and the change in color in the medians is comparable to these uncertainties.

It is interesting to compare our measured *HST*/ACS color profile to that found from Subaru photometry in Arnold et al. (2011), which extends roughly twice as far in radius from NGC 3115 but with reduced precision. We recover the jump in color in the red subpopulation located around $R_g \sim 6$ kpc seen in Arnold et al. (2011), corroborating the visible features in the red color gradient we see in our data. Arnold et al. (2011) found gradients of $-0.17 \pm 0.04$ dex dex$^{-1}$ and $-0.38 \pm 0.06$ dex dex$^{-1}$ for the red and blue GCs, respectively. Thus, our color gradients agree to within the uncertainties on the quantities, although there is scatter. It is worth noting that the Arnold et al. (2011) values must also use an empirical correction from $(g-i)$ to $(g-z)$ to compare metallicities (equation A1 from Usher et al. 2012); the uncertainties on this filter conversion are also neglected in the Arnold et al. (2011) values. The photometric data from Arnold et al. (2011) extended roughly twice as far in radial distance from the center of NGC 3115 revealing, further structure in the red subpopulation including a large decrease in color around 15 kpc and a flattening of the color gradient farther out. It is likely that these two effects cancel out somewhat in the measured gradient from Arnold et al. (2011), leading to comparable values from both studies.

We also consider the Gaussian kernel density color distribution as a function of $R_g$. In Fig. 11 we plot the Gaussian kernel distribution evaluated at five 2.8 kpc distance bins. The blue peak is located fairly consistently and, with the exception of the second farthest bin, decreases in color monotonically. On the other hand, the red peak again displays visible color substructure. The peak moves around without a clear trend and indeed disappears completely in the second farthest bin, even though the overall subpopulation still displays a color gradient.

Generally speaking, declining galaxy metallicity gradients with a flattening at large galactocentric radii are a standard prediction of two-phase assembly scenarios (Naab et al. 2009; Bezanson et al. 2009; Hirschmann et al. 2013). An early monolithic collapse produces the inner gradient, while the outer flattening is produced by repeated accretion events. The existence of the outer flattening is unclear in our data, probably due to the limited radial coverage provided by the ACS mosaic. Smaller accreted satellites should be metal-poor in comparison to the central galaxy. Given the consistency of our measured profiles to those inferred in Arnold et al. (2011) from Suprime-Cam and spectroscopic data, we refer the reader to that paper for a more in-depth discussion of this point in the context of NGC 3115, including the incorporation of velocity information both in the GC population and integrated starlight.

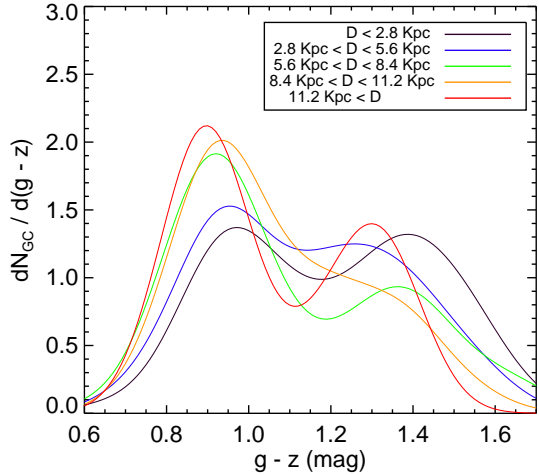

**Figure 11.** Gaussian kernel density distributions of color for the GC subpopulation in a series of 2.8 kpc projected galactocentric radius bins. The blue peak is fairly consistent in location, and decreases in color monotonically with the exception of the second farthest bin. In addition, the strength of the blue peak consistently increases with distance from NGC 3115. The red peak, on the other hand, displays possible color substructure. There is no consistent trend with galactocentric radius, and indeed the red peak disappears entirely in the second farthest radial bin.

### 3.4. *Trends in Size*

In Fig. 12, we plot the measured $R_h$ as a function of projected galactocentric distance from the center of NGC 3115. Measurements of $R_h$ tend to show a large variance. Many clusters of both subpopulations display $R_h$ values significantly larger than those common for clusters, which are close to $R_h \sim 2-4$ pc. As we are chiefly interested in the trends of the most-likely GCs, we remove these outliers by only including GCs with $R_h < 8$ pc in this subsection, as explained in §2.3. In general, inclusion of larger GCs does not affect the overall trend of $R_h$ with distance, but the uncertainties on the medians and gradients increase.

It is well documented in the literature that $R_h$ values tend to increase with galactocentric distance, and that the blue GCs tend to have larger sizes than red GCs, with differences typically on the order of $\sim 20\%$ (Kundu & Whitmore 1998; Larsen et al. 2001; Jordán et al. 2005). We find median sizes of $2.25^{+0.10}_{-0.04}$ pc for the blue subpopulation and $2.06^{+0.11}_{-0.14}$ pc for the red subpopulation. 68% confidence intervals are estimated through bootstrapping. These median values correspond to a fractional difference of $\sim 10\%$. This fraction is somewhat low compared to the above literature values, but the uncertainties on the medians and scatter of the measurements is large.

Trends in $R_h$ with radius are typically parameterized in the form $R_h = a(R/R_e)^b$. Performing a least squares fit to our data, we find the following relations:

$$R_{h,blue} = (0.31 \pm 0.02)(R_g/R_e)^{0.14 \pm 0.05} \text{ pc} \quad (10)$$

for the blue GCs and

$$R_{h,red} = (0.30 \pm 0.02)(R_g/R_e)^{0.08 \pm 0.07} \text{ pc} \quad (11)$$

for the red GCs. We plot these best fitting relations on Fig 12. In the literature, as with the median sizes of the respective subpopulations, differences are typically observed in the power-law slopes of the best-fitting $R_h$ vs $R/R_e$ relations, with blue GCs typically featuring smaller slopes than the red GCs. Our slopes are broadly consistent with those found in the literature (Spitler et al. 2006; Gómez & Woodley 2007; Harris & Zaritsky 2009; Harris et al. 2010; Strader et al. 2012), although literature studies have typically found the slopes of the red and blue subpopulations to be distinct. Instead, we find power-law slopes to be fairly consistent between the two subpopulations.

For completeness, we also performed a simple linear least squares fit to the data. We find best fit relations of the form

$$R_{h,blue} = [(0.17 \pm 0.09)R_g/R_e + (2.0 \pm 0.2)] \text{ pc} \quad (12)$$

for the blue clusters and

$$R_{h,bed} = [(0.11 \pm 0.07)R_g/R_e + (2.0 \pm 0.2)] \text{ pc} \quad (13)$$

for the red clusters. Thus, while the strengths of the gradients are roughly consistent, the significance of the red gradient is much lower than that of the blue.

We also plot the median values of each subpopulation in 4 radial bins of $\sim 3$ kpc. 68% uncertainties on the medians from bootstrapping are included on the median values. We bin in radius instead of number here because we are interested in comparing the properties of the two subpopulations at the same distances, as opposed to looking at internal gradients. For the blue subpopulation, the median $R_h$ values increase monotonically with radius out to the final bin. Uncertainties on the values tend to be small, and the points are well clustered around the best-fitting relation. The red subpopulation, on the other hand, displays significant scatter around the best-fitting line, especially at large galactocentric distances, where there are very few clusters. It is questionable whether the red subpopulation displays any increase at all, given the few clusters located at large $R_g$ and the large scatter of the subpopulation.

The small difference in median cluster size and the lack of a distinction in power-law slopes is somewhat interesting. The color-bimodality is especially pronounced in the GC system of NGC 3115, and it is well established that this bimodality also corresponds to a metallicity bimodality (Brodie et al. 2011). However, despite the distinct differences in color and metallicity of the subpopulations, they are not nearly as distinct in their sizes.

There is discussion in the literature about whether the observed size difference between red and blue clusters is due to an intrinsic, metallicity-dependent process (e.g., Jordán 2004a, or a projection effect due to the fact that red clusters tend to be more centrally concentrated (e.g., Larsen & Brodie 2003). The main prediction of the projection explanation is that the difference between the red and blue clusters will disappear at large radii, while the intrinsic explanation predicts that the separation will remain at all radii. While inferring broad trends in the data is questionable given the scatter, we do not see any evidence for a decreasing difference between red and blue $R_h$ at larger $R_g$. This observation would seem to favor the intrinsic, metallicity-dependent explanation. However, this inference is tenuous, and the fact that the power-law

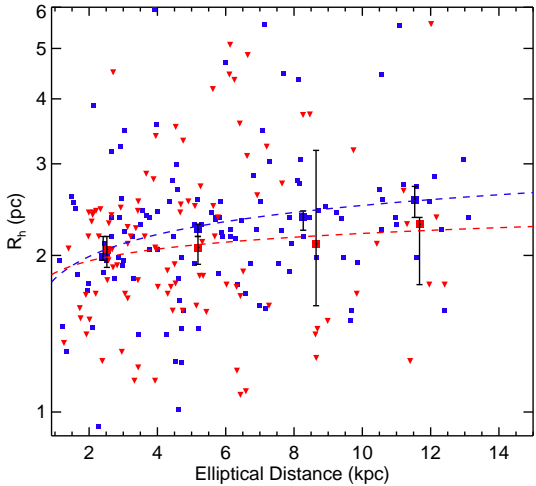

**Figure 12.** Plot of cluster half-light radius vs. projected galactocentric distance from NGC 3115. We also plot median half-light radii for equal number bins of clusters, measured separately for both subpopulations. 68% uncertainties on the medians from bootstrapping are also plotted. $R_h$ measurements for both subpopulations of GCs become larger with increasing distance from NGC 3115. In general, the blue subpopulation displays slightly larger half-light radii than the red subpopulation.

slopes and median sizes of the two subpopulations are so similar is also peculiar in this context.

Observational support has been found for both the intrinsic explanation (Harris 2009; Paolillo et al. 2011; Blom et al. 2012; Strader et al. 2012) and the projection hypothesis (Spitler et al. 2006). It is worth emphasizing that those studies that have found support for the intrinsic metallicity explanation have generally focused on giant elliptical galaxies. However, Spitler et al. (2006) examined M104, an edge-on SA galaxy in a group environment. Interestingly, M104 also displayed an unusually small difference (14%) in median values between the red and blue GC subpopulations, consistent with those we have measured for NGC 3115. The environmental and morphological conditions of M104 are quite similar to NGC 3115. Further investigation of GC $R_h$ measurements across a wider range of morphological and environmental properties will serve to shed light on the full importance of cluster vs. group/field environments, morphology, and inclination in the size differences between the red and blue GC subpopulations. Given the similar trends in $R_h$ with $R_g$ and the consistency in the median size difference between this work and Spitler et al. (2006), we suppose that inclination and morphology in particular may play a role in the relative importance of projection effects compared to intrinsic size differences. In the end, there is only a limited amount we may infer about the NGC 3115 GC system other than that the two subpopulations are not very distinct in their sizes.

### 3.5. *Ultra Compact Dwarf Candidates and other Extended Objects*

There is significant discussion in the literature regarding the dividing line between the largest star clusters and the most compact galaxies. This division is typically explored in the parameter space of $R_h$ and absolute magnitude. In the past, observational searches of this parameter space have revealed ultra compact dwarfs (UCDs), which are brighter and more extended than typical GCs (Hilker et al. 1999; Drinkwater et al. 2000; Phillipps et al. 2001). They occupy a middle ground in size between GCs and dwarf elliptical galaxies. A traditional view of UCDs was that they represent the continuation of the size-luminosity trend to larger and brighter clusters, a view motivated by the fact that the first UCDs to be discovered were naturally the brightest. An apparent luminosity gap was also observed between UCDs and other extended objects, such as Local Group extended clusters (ECs) (e.g., Huxor et al. 2005), extragalactic "faint fuzzies" (FFs) (e.g., Brodie & Larsen 2002), and "diffuse star clusters" (e.g., Peng et al. 2006b). No universal definition of a UCD exists, but a "traditional" definition of a UCD might be an object with luminosity $\sim 10^7 \, L_\odot$ ($M_z \sim -13$) and $R_h \sim 20$ pc, with significant scatter around these values.

However, as optical studies have continually probed lower luminosities, the boundaries of the UCD population have become somewhat ill-defined. Brodie et al. (2011) adopted provisional criteria of $M_V < -8.5$ and $R_h > 10$ pc to account for the presence of new spectroscopically confirmed objects around M87 at lower luminosities. Forbes et al. (2013) noted a number of new spectroscopically confirmed objects which fully bridge the gap between the UCD and FF populations. In light of recent observational data, it is unclear that performing any luminosity cut on the extended object population is observationally motivated. There may still be interesting theoretical motivations, such as the $2 \times 10^6 \, M_\odot$ boundary for self enrichment estimated by Bailin & Harris (2009) or relaxation timescale arguments such as that presented in Misgeld & Hilker (2011). However, for our brief consideration in this paper, we ultimately choose to employ no selection criteria aside from a size cut of $R_h > 8$ pc. We wish to include all interesting extended objects, including traditional UCDs, FFs, and intermediate transition objects, in a catalog for potential future spectroscopic follow-up, and the proximity of NGC 3115 offers an ideal target for studying precisely these sorts of objects. Throughout this section we will frequently refer to all of these objects as "UCD candidates," despite the fact that many are not in traditional UCD parameter space and are instead more similar to FFs and other extended objects.

UCDs have primarily been studied in cluster environments, where targets tend to be more dense. Mieske et al. (2004) examined UCD-like objects in Fornax and found their formation was consistent both with a stripping scenario and a cluster merger scenario. Haşegan et al. (2005) and Price et al. (2009) studied UCD-like objects in Virgo and Coma, respectively, using *HST*/ACS imaging and found several strong candidates in each. Their results favored tidal stripping scenarios based on extrapolations of scaling relations. However, other studies have found cluster merger scenarios to be more plausible scenarios for UCD formation (e.g., Mieske et al. 2006a; Kissler-Patig et al. 2006). There is increasing evidence that multiple scenarios are necessary to fully explain the observed properties of UCDs across multiple environments (Taylor et al. 2010; Norris & Kannappan 2011; Da Rocha et al. 2011; Penny et al. 2012).

It is unclear how environmental effects will affect the formation of UCD objects. If the stripping of galaxies is the primary mechanism, then naturally these objects will be more likely to form in denser environments. Only a few studies have investigated the properties of UCDs found in group/field environments (e.g., Evstigneeva et al. 2007; Hau et al. 2009; Norris & Kannappan 2011; Da Rocha et al. 2011, Norris et al. 2014 (MNRAS, Submitted)). Given the potential environmental dependence, the detection of candidates in non-cluster environments such as NGC 3115 is of value for future inves-tigations of UCD formation.

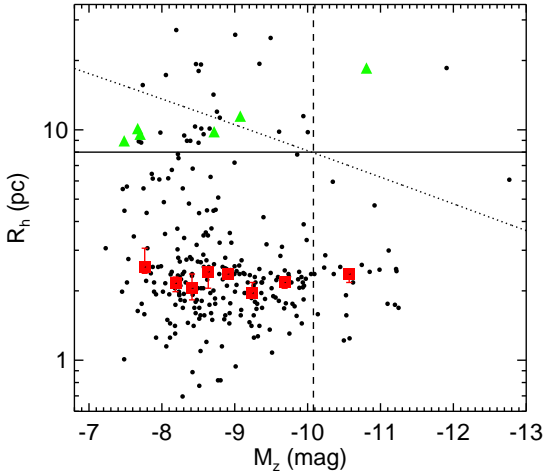

**Figure 13.** Plot of measured GC half-light radii against absolute $z$ magnitude, $M_z$. We identify all clusters with $R_h > 8.0$ pc as potential UCD candidates. Six candidates have measured radial velocities from Arnold et al. (2011) that confirm NGC 3115 membership; we highlight these as green triangles. We recover a number of candidates that have luminosities consistent with faint GCs, but larger sizes. We also plot the median measured $R_h$ values for 8 equal-number bins of clusters for only those clusters with $R_h < 8.0$ pc, with 68% uncertainties on these medians from bootstrapping included. The vertical line represents a $10^6 \,M_\odot$ cut, the horizontal line represents the adopted $R_h < 0.8$ pc size cut, and the diagonal line represents the dividing line for objects which will have undergone significant dynamical evolution within one Hubble time.

As our ultimate goal in this section is inclusion of as many candidate UCDs as possible, we relax the size restrictions used in creating our catalog and no longer employ any maximum size cuts. We still include the depth cuts employed for our measured $R_h$ sizes by only considering candidates with signal to noise greater than 50.

In Fig. 13, we plot the measured $R_h$ against the absolute magnitude of clusters. Traditionally, UCDs are thought of as occupying the parameter space of bright clusters with $M_z < -11$. However, as we are specifically interested in investigating faint candidates, we consider everything with measured $R_h > 8$ pc a UCD candidate. We identify 31 such candidates, which we list in Table 3. Six of the brightest candidates have measured DEIMOS radial velocities from Arnold et al. (2011) consistent with NGC 3115 membership. We plot these with solid green triangles.

We include three potential definitions for what one might consider a UCD. The horizontal line is the size cut we adopt for our table of UCD candidates, $R_h = 8$ pc. The vertical line is a luminosity cut which corresponds to a mass of $M = 10^6 \,M_\odot$ ($\sim 7 \times 10^5 \,L_\odot$). Finally the diagonal line is the size–mass relation from Misgeld & Hilker (2011) (adapted from Dabringhausen et al. 2008) for an assumed relaxation timescale. Objects below the line will have been able to undergo significant dynamical evolution over a Hubble time, while those above would be expected to show a mass-size relation.

We identify a number of fainter UCD candidates, which would populate the parameter space occupied by the faint UCD sequence parallel to the GC population. However, it is unclear how distinct this population is from the normal GC sequence in our data. These candidates are also all towards the fainter end of the GC sequence, where background galaxy contamination will become most prevalent. Indeed, objects in this region of parameter space are in the regions where we would have expected clusters to undergo significant dynamical evolution, making it more likely that sources here are contaminants. The fact that many have red colors (see Fig. 7) reinforces this caveat. Further spectroscopic follow-up of UCD candidates will allow a more thorough investigation of this parameter space through rejection of background galaxy contamination.

In addition, we note that several of these objects have significant discrepancies between ground-based Suprime-Cam imaging and ACS photometry, as large as a couple magnitudes for a few objects. Such an offset is likely due to an object having a size-dependent correction applied which is either too large (i.e. the object is actually significantly smaller than what we measure using our King profile fit and therefore we overestimate the correction) or too small (the opposite case). For these large, extended objects, the S/N towards the edges of the profile is small, and so the uncertainties on measured sizes can be large. In addition, for any object which is in fact a background galaxy, the King profile is likely a poor parameterization of the light profile to begin with. For the reason of background contamination, we caution against inferring too much from non-spectroscopically confirmed candidates.

Six candidates, UCD1, UCD5, UCD 8, UCD15, UCD16, and UCD23, are spectroscopically confirmed to have NGC 3115 membership. While UCD1 occupies traditional UCD parameter space, the remainder extend to the fainter luminosities typically occupied by faint fuzzies. These additional confirmed candidates do support an interpretation that a similar population of extended clusters exists parallel to a traditional GC sequence, perhaps the result of a distinct formation mechanism. However, more spectroscopically confirmed clusters are necessary to evaluate the full extent of this population; some significant number of the unconfirmed candidates may still be contaminants.

In addition, object A1, not in the UCD catalog, is noteworthy for its density. Its size is large for, but not inconsistent with, a GC ($R_h \sim 6$ pc). However, it also features an extremely high luminosity, with $M_z \sim -12.7$, corresponding to a mass of $1.18 \times 10^7 \,M_\odot$. This object is located very close to the center of NGC 3115, perhaps indicating some systematic effect from the galaxy light on the *ishape* measurement.

We also plot median $R_h$ values for eight equal-numbered magnitude bins. We do not see a strong trend in $R_h$ with absolute magnitude. Other studies have typically found correlations between $R_h$ and absolute magnitude for the brightest clusters. We do find a monotonic increase in $R_h$ with increasing luminosity for $M_z < -9$. However, the median values below this point display significant scatter. The hints of a trend we see do match that found in other studies, wherein the $R_h$ of a typical cluster appears to be larger both for the faintest and brightest candidates, and smallest for those of intermediate luminosity (Haşegan et al. 2005; Evstigneeva et al. 2008; Dabringhausen et al. 2008; Harris 2009; Harris et al. 2010; Misgeld & Hilker 2011; Strader et al. 2012). However, the trend is very weak in our data.

## 4. X-RAY/GC MATCHING

In this section, we discuss those GCs identified in §2.7 which are coincident with X-ray sources. We identify 45 X-ray sources within 3-$\sigma$ X-ray PSF distances from our GC catalog. For those sources in the ACS FOV, we adopt the ACS ($g-z$) color division between the red and blue subpop-

Table 3
UCD Candidates in ACS Data

| ID | R.A. (J2000) (Deg.) | Dec. (J2000) (Deg.) | $M_g$ (mag) | err (mag) | $M_z$ (mag) | err (mag) | $R_h$ (pc) | Vel.[A] (km s$^{-1}$) | X-ray?[B] (Y/N) |
|---|---|---|---|---|---|---|---|---|---|
| UCD1  | 151.30216 | -7.7355695 | -10.748 | 0.002 | -11.847 | 0.002 | 18.55 | 393 | N |
| UCD2  | 151.22176 | -7.7326969 | -9.070  | 0.012 | -10.602 | 0.008 | 80.16 | -   | N |
| UCD3  | 151.34482 | -7.7344961 | -8.947  | 0.011 | -9.964  | 0.011 | 70.63 | -   | N |
| UCD4  | 151.31997 | -7.7700997 | -8.710  | 0.005 | -9.943  | 0.004 | 9.77  | -   | N |
| UCD5  | 151.28906 | -7.7722947 | -9.016  | 0.004 | -9.881  | 0.004 | 11.47 | 518 | N |
| UCD6  | 151.28735 | -7.773133  | -8.925  | 0.011 | -9.780  | 0.013 | 70.63 | -   | N |
| UCD7  | 151.37419 | -7.6955633 | -8.412  | 0.015 | -9.745  | 0.012 | 73.62 | -   | N |
| UCD8  | 151.31825 | -7.7340325 | -8.657  | 0.005 | -9.549  | 0.005 | 9.81  | 844 | N |
| UCD9  | 151.29046 | -7.7393686 | -8.197  | 0.011 | -9.432  | 0.012 | 25.05 | -   | N |
| UCD10 | 151.34388 | -7.7470892 | -7.853  | 0.017 | -9.279  | 0.012 | 19.35 | -   | N |
| UCD11 | 151.27333 | -7.7342231 | -7.483  | 0.015 | -8.947  | 0.011 | 25.86 | -   | N |
| UCD12 | 151.25691 | -7.6970153 | -7.531  | 0.009 | -8.735  | 0.007 | 11.29 | -   | N |
| UCD13 | 151.31608 | -7.6916319 | -7.167  | 0.013 | -8.696  | 0.011 | 11.99 | -   | N |
| UCD14 | 151.32590 | -7.7242011 | -7.337  | 0.014 | -8.648  | 0.012 | 14.22 | -   | N |
| UCD15 | 151.28940 | -7.6638911 | -7.608  | 0.009 | -8.595  | 0.008 | 10.13 | 878 | N |
| UCD16 | 151.36446 | -7.6930417 | -7.642  | 0.009 | -8.513  | 0.009 | 9.57  | 661 | N |
| UCD17 | 151.36931 | -7.6728572 | -7.601  | 0.010 | -8.481  | 0.012 | 10.13 | -   | N |
| UCD18 | 151.25418 | -7.7672714 | -7.177  | 0.015 | -8.477  | 0.012 | 19.21 | -   | N |
| UCD19 | 151.30366 | -7.6752959 | -7.426  | 0.012 | -8.445  | 0.012 | 18.01 | -   | N |
| UCD20 | 151.32090 | -7.7312989 | -7.209  | 0.015 | -8.444  | 0.014 | 8.79  | -   | Y |
| UCD21 | 151.25392 | -7.7509972 | -7.024  | 0.018 | -8.408  | 0.014 | 19.3  | -   | N |
| UCD22 | 151.29983 | -7.7638275 | -7.221  | 0.011 | -8.395  | 0.01  | 10.34 | -   | N |
| UCD23 | 151.34029 | -7.6925642 | -7.423  | 0.011 | -8.330  | 0.013 | 8.96  | 579 | N |
| UCD24 | 151.29070 | -7.8077822 | -7.130  | 0.014 | -8.284  | 0.011 | 8.96  | -   | N |
| UCD25 | 151.29154 | -7.7086903 | -7.063  | 0.016 | -8.247  | 0.015 | 9.45  | -   | N |
| UCD26 | 151.32115 | -7.7692714 | -7.309  | 0.018 | -8.137  | 0.019 | 27.11 | -   | N |
| UCD27 | 151.36293 | -7.6982394 | -6.992  | 0.019 | -7.996  | 0.017 | 17.30 | -   | N |
| UCD28 | 151.28682 | -7.7838389 | -6.457  | 0.020 | -7.920  | 0.014 | 9.73  | -   | N |
| UCD29 | 151.24861 | -7.704778  | -7.025  | 0.015 | -7.678  | 0.018 | 15.64 | -   | N |
| UCD30 | 151.28885 | -7.7637978 | -6.828  | 0.015 | -7.656  | 0.019 | 8.80  | -   | N |
| UCD31 | 151.23764 | -7.7306361 | -6.444  | 0.018 | -7.623  | 0.016 | 8.91  | -   | N |

**A**: Heliocentric velocity, if available, from the Potá et al. (2013) catalog.
**B**: Whether or not the GC candidate hosts an X-ray source.

ulations. Otherwise, we adopt the Supime-Cam photometry $(g-i)_0$ color division. We color-code and plot the two distributions in Fig 14. In general, the red X-ray GCs appear to be more centrally concentrated than the blue X-ray GCs, matching the trend in the population of GCs without X-ray sources.

Of the 45 sources, 29 are associated with red GCs, while 16 are associated with blue GCs. Given a total of 291 unique red and 490 unique blue GC candidates across both optical catalogs, we find a $\sim$10% chance that a red GC hosts an X-ray source and a $\sim$3% chance that a blue GC hosts an X-ray source. These fractions are broadly consistent with those reported in the literature (Jordán 2004b; Fabbiano 2006; Kim et al. 2006; Sivakoff et al. 2007) although they are certainly on the high side for both values (typically $\sim$5% for red GCs and $\sim$2% for blue GCs). If we match to only the more reliable ACS sources, the fraction of each increases to $\sim$14% for red GCs and to $\sim$7% for blue GCs, both extremely high for X-ray/GC coincidence. If we restrict our analysis to just those GCs which are ACS imaged and spectroscopically confirmed, rates increase further to $\sim$14% for both the red and blue subpopulations. Note, however, that the spectroscopically confirmed sample is naturally biased to the brightest clusters due to observational constraints. Brighter clusters are expected to be more likely to host LMXBs (Sivakoff et al. 2007). It is likely that these higher rates of X-ray/GC coincidence are due to the depth of the X-ray observations compared to similar studies, as well as the smaller distance of NGC 3115 compared to most other early-type galaxy GC studies.

In Fig. 14, we also note one particular feature in the blue matched GC subpopulation, a linear structure of X-ray associated GCs located at RA $\sim$ 151.31, Dec $\sim -7.65$. It is unclear exactly how to interpret this organized structure of X-ray GCs. They are all detected in the ACS mosaic, but apart from this they do not have any truly distinguishing features. They have typical sizes, colors, and luminosities typical of blue GCs. Three of the GCs in the line have measured radial velocities, all in the same direction as the rotation of NGC 3115's disk. The existence both of the plane of blue GCs and the overabundance of X-ray sources in the blue GCs may be linked, but given our current analysis, such an inference is speculative, with unclear implications.

We do not discuss further the properties of the X-ray hosting GCs beyond initial rates of incidence, nor the properties of the X-ray sources themselves. Analyses of these properties will be performed in a subsequent paper (Lin, D. et al 2014, in preparation).

5. SUMMARY

We have performed photometry and size measurements for 360 GC candidates in *HST*/ACS imaging of NGC 3115. We have also presented Suprime-Cam photometry for 421 additional candidates. The bimodality of the system is very obvious in our data. There is evidence for a blue-tilt in the blue subpopulation, and we see weak hints of an opposing trend in the red GCs. Both subpopulations display a color gradient as a function of distance from the center of NGC 3115, and the magnitude of the gradient is similar to that found at larger radii in Arnold et al. (2011). The blue GCs display mono-tonic behavior, decreasing in color uniformly. However, the red GCs display visible color substructure in addition to the overall gradient. The size behavior of the two subpopulations is somewhat unusual. While the blue GCs are larger on average than the red GCs, the ratio of the average sizes is closer to unity than is typical for a GC system. We are unable to clearly confirm either a projection or an intrinsic explanation for the size distinction, but given the similarity of our result to that found in Spitler et al. (2006) for M104, we suppose that the morphology and inclination of the galaxy may have a significant effect on the measured relative sizes of the two subpopulations.

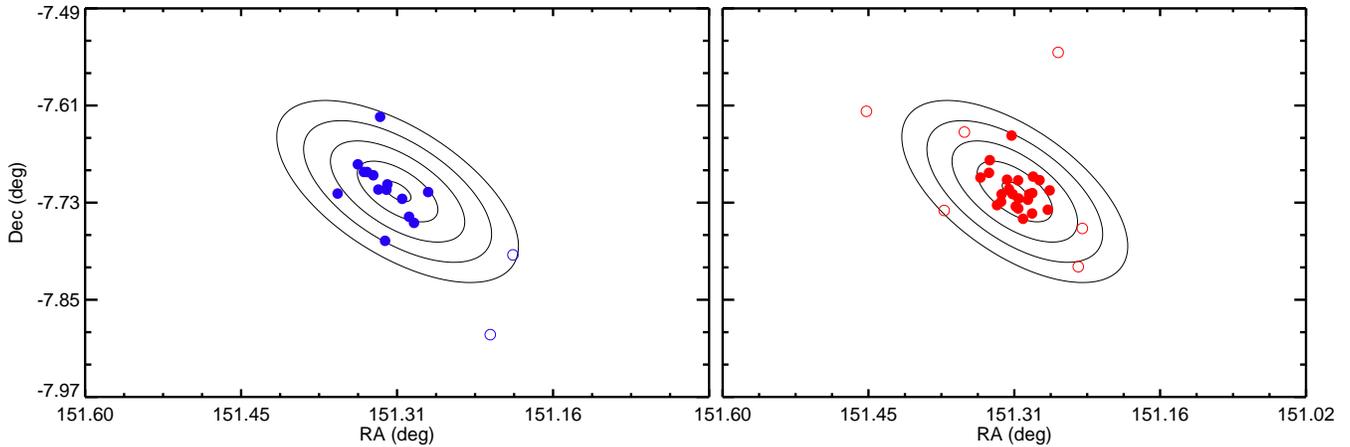

**Figure 14.** Spatial distribution of blue (left panel) and red (right panel) GCs with X-ray matches. Closed circles are detected in the ACS catalog, while open circles are only detected in the large FOV Suprime-Cam imaging. Contours of 1, 3, 5, 7, and 9 $R_e$ are plotted for reference.

We identify 31 candidate UCD objects, including six with spectroscopic confirmation. Given their colors, it is possible that many candidates without measured velocities are in fact background contaminants. In addition, after matching our ACS sources with companion X-ray data, we find 29 X-ray sources associated with red GCs and 16 with blue GCs. The fraction of X-ray hosting GCs is larger for both subpopulations than is typical in the literature, especially for blue GCs, likely due to the increased depth of the X-ray data. We also observe an interesting linear spatial distribution in the blue X-ray hosting population. The implications of this distribution and its link to the overabundance of X-ray sources in blue GCs are unclear.

We thank D. Forbes and K. Woodley for helpful comments in the course of this work. This research is based on observations made with the NASA/ESA *Hubble Space Telescope*, obtained from the data archive at the Space Telescope Science Institute. STScI is operated by the Association of Universities for Research in Astronomy, Inc. under NASA contract NAS 5-26555. ZGJ is supported in part by a National Science Foundation Graduate Research Fellowship. Based in part on data collected at Subaru Telescope (operated by the National Astronomical Observatory of Japan) via a Gemini Observatory time exchange (GN-2008A-C-12), and also the 6.5 m Magellan Telescopes located at Las Campanas Observatory, Chile. Some of the data presented herein were obtained at the W. M. Keck Observatory, operated as a scientific partnership among the California Institute of Technology, the University of California and the National Aeronautics and Space Administration, and made possible by the generous financial support of the W. M. Keck Foundation. This material is based upon work supported in part by the National Science Foundation under Grants AST-1211995 and AST-1109878. This material is based upon work supported in part by HST-GO-12759.12-A. GRS acknowledges the support of an NSERC Discovery Grant.